\documentclass[twocolumn,showpacs,preprintnumbers,amsmath,amssymb]{revtex4}
\usepackage{graphicx}
\usepackage{dcolumn}
\usepackage{bm}
\newcommand{\bea}{\begin{eqnarray}}
\newcommand{\eea}{\end{eqnarray}}
\newcommand{\be}{\begin{equation}}
\newcommand{\ee}{\end{equation}}
\newcounter{saveeqn}
\newcommand{\alpheqn}{\setcounter{saveeqn}{\value{equation}}%
\setcounter{equation}{0}%
\renewcommand{\theequation}{\mbox{\arabic{saveeqn}\alph{equation}}}}
\newcommand{\reseteqn}{\setcounter{equation}{\value{saveeqn}}%
\renewcommand{\theequation}{\arabic{equation}}}
\newcommand{\xbf}[1]{\mbox{\boldmath $ #1 $}}
\def\shiftdown#1{#1\llap{\lower.04ex\hbox{#1}}}

\begin{document}
\title{{\Large Baryon octupole moments}}
\thanks{published in: Eur. Phys. J. A {\bf 35}, 267 (2008).}


\author{A. J. Buchmann$^{1}$} 
\email{alfons.buchmann@uni-tuebingen.de}
\author{E. M. Henley$^{2}$}  
\email{henley@u.washington.edu}

\affiliation{$^1$Institut f\"ur Theoretische Physik, Universit\"at 
T\"ubingen\\
Auf der Morgenstelle 14, D-72076 T\"ubingen, Germany \\
$^2$Department of Physics and Institute for Nuclear Theory, 
Box 351560, \\
University of Washington, Seattle, WA 98195, U.S.A.}

\begin{abstract}
We report on a calculation of higher electromagnetic 
multipole moments of baryons in a non-covariant quark model approach.
The employed method is based on the underlying spin-flavor symmetry
of the strong interaction and its breaking.
We present results on magnetic octupole moments of decuplet
baryons and discuss their implications.
\end{abstract}
\maketitle
\section{Introduction}
\label{intro}

Electromagnetic multipole moments of baryons are interesting observables. 
They are directly connected with the spatial charge 
and current distributions in baryons, and thus contain fundamental  
information about their internal structure and geometric shape. 

For example, recent electron-proton and photon-proton scattering 
experiments, exciting the lowest lying nucleon resonance $\Delta^+(1232)$ 
have provided evidence for a nonzero quadrupole moment 
$Q_{p \to \Delta^+}\!\! \approx \!\! -0.08$ fm$^2$~\cite{Ber03,Tia03,Bla01}
and hence for a
nonspherical charge distribution in baryons. The measured sign and
magnitude are in agreement with the quark model prediction 
$Q_{p \to \Delta^+}= \, r_n^2/\sqrt{2}$, where $r_n^2$ 
is the neutron charge radius~\cite{Buc97}.
It has been suggested that a transition quadrupole moment of 
this sign arises because the proton has a prolate and the $\Delta^+$ an 
oblate charge distribution~\cite{Hen01}. 
For a recent review see Ref.~\cite{Pas07}. 

While there is a large body of literature on baryon magnetic dipole moments, 
relatively little is known about the next higher multipole moments, 
that is the magnetic octupole moments $\Omega$ 
of decuplet baryons~\cite{But94}.
Presently, we neither know the sign nor the size of 
these moments. This information is needed to reveal   
further details of the current distribution in baryons beyond 
those available from the magnetic dipole moment~\cite{kot02}. 

The purpose of this paper is to calculate the
magnetic octupole moments of baryons and to draw some  
conclusions concerning the shape of their spatial current distributions.
\vspace{-0.2cm}

\section{Method}
\label{sec:1}
We use a general parametrization (GP) 
method developed by Morpurgo~\cite{Mor89},
which incorporates SU(6) symmetry and its breaking similar to the 
$1/N_c$ expansion. 
The basic idea is to {\it formally} define, for the observable at
hand, a QCD operator $\Omega$ and QCD eigenstates $\vert B \rangle$
expressed explicitly in terms of quarks and gluons. The corresponding matrix 
elements can, with the help of the unitary operator $V$, be reduced to an
evaluation in a basis of pure three-quark states
$\vert\Phi_B \rangle $ with orbital angular momentum
$L=0$
\begin{equation}
\label{map}
\left \langle B \vert \Omega \vert B \right \rangle =
\left \langle \Phi_B \vert
V^{\dagger}\Omega V \vert \Phi_B \right \rangle =
\left \langle W_B \vert
{ {\tilde \Omega}} \vert W_B \right \rangle \, .
\end{equation}
The spin-flavor wave functions
contained in $\vert \Phi_B \rangle$ are denoted by $\vert W_B\rangle $.
The operator $V$ dresses the pure three-quark
states with $q\bar q$ components and gluons and
thereby generates
the exact QCD eigenstates $\vert B \rangle $. 

One then writes the most general expression for the operator
${ {\tilde \Omega}}$ that is
compatible with the space-time and inner QCD symmetries.
Generally, this is a sum of one-, two-, and three-quark
operators in spin-flavor space multiplied by {\it a priori} unknown constants
which parametrize the orbital and color space
matrix elements.
Empirically, a hierarchy in the importance
of one-, two-, and three-quark operators is found.
This fact can be understood
in the $1/N_c$ expansion where
two- and three-quark operators describing second and third
order SU(6) symmetry breaking
are usually suppressed by powers of $1/N_c$ and $1/N_c^2$ 
respectively compared to one-quark operators associated with first order 
symmetry breaking~\cite{Ncpapers}.
The GP method has recently been used to calculate 
charge radii and quadrupole moments of baryons~\cite{Mor99,Hen02}.
\vspace{-0.2cm}

\section{Calculation}
The magnetic octupole moment operator $\Omega$ usually given
in units $[{\rm fm}^2 \, \mu_N]$ and
normalized as in Ref.~\cite{Don84} can be written as
\bea 
\label{M1andM3}
\Omega & = & \frac{3}{8}
\int \! dr^3 (3 z^2-r^2) \, ({\bf r} \times {\bf J}({\bf r}))_z,
\eea
where ${\bf J}({\bf r})$ is the spatial current density and 
$\mu_N$ the nuclear magneton. 
This definition is analogous to the one for the
charge quadrupole moment~\cite{Hen01} if the magnetic moment density 
$({\bf r} \times {\bf J}({\bf r}))_z$ is replaced
by the charge density $\rho({\bf r})$. 
Thus, the magnetic octupole moment measures 
the deviation of the spatial magnetic moment distribution from
spherical symmetry. More specifically, for a prolate (cigar-shaped) 
magnetic moment distribution $\Omega >0$, 
while for an oblate (pancake-shaped) magnetic moment distribution 
$\Omega <0$. We also see from Eq.(\ref{M1andM3}) that the typical size of a 
magnetic octupole moment is 
\be
\Omega \simeq r^2 \, \mu  
\ee
where $\mu$ is the magnetic moment and $r^2$ a size parameter 
related to the quadrupole moment of the system.

To construct an octupole moment operator ${\tilde \Omega}$ 
in spin-flavor space along the lines outlined in sect. 2 
we need a tensor of rank 3 in spin space. Clearly, this operator
must involve the Pauli spin matrices of three different 
quarks~\cite{comment0} and can be built in two different ways.
First, we can construct it from a three-body quadrupole 
moment operator multiplied by the spin of the third quark, 
\begin{eqnarray} 
\label{para2}
{\tilde \Omega}_{[3]} & = & C \sum_{i \ne j \ne k }^3 e_k  
\left ( 3 \sigma_{i \, z} \sigma_{ j \, z} - 
\xbf{ \sigma}_i \cdot \xbf{ \sigma}_j \right )\xbf{ \sigma}_{k},
\end{eqnarray} 
where $C$ is a constant and 
$e_k=(1 + 3 \tau_{k \, z})/6$ is the charge of the k-th quark.
The $z$-component
of the Pauli spin (isospin) matrix $\xbf{ \sigma}_i$ ($\xbf{\tau}_i$) 
is  denoted by $\sigma_{i \, z}$ ($\tau_{i \, z}$).
Second, we can build it from a two-body quadrupole moment operator 
by replacing  $e_k$ by $e_i$ in Eq.(\ref{para2}). Thus, 
it appears that there are two different operator structures and 
corresponding GP constants to be determined from experiment. 
However, from the point 
of view of broken SU(6) spin-flavor symmetry~\cite{Gur64}, 
there is a unique three-body operator containing a rank 3 
spin tensor~\cite{comment2}.

The magnetic octupole moments $\Omega_{B^*}$ 
are then obtained by sandwiching the operator in Eq.(\ref{para2}) 
between the three-quark spin-flavor wave functions $\vert W_{B^{*}} \rangle $.
For example, for $\Delta(1232)$ baryons one obtains
\bea
\label{twothree}
\Omega_{\Delta} & = &\langle W_{\Delta} 
\vert {\tilde {\Omega}}_{[3]} 
\vert  W_{\Delta} \rangle  =  4 \, C \, q_{\Delta},
\eea
where $q_{\Delta}$ is the $\Delta$ charge.
Similarly, the magnetic octupole moments for the other decuplet baryons
are calculated.
In this way Morpurgo's method yields an efficient parameterization
of baryon octupole moments in terms of just one unknown parameter $C$.
\vspace{-0.2cm}

\section{Results} 
\label{sec:results}
In the second column of Table~\ref{octumom} 
we show our results for the decuplet octupole moments 
expressed in terms of the GP constant $C$ 
assuming that SU(3)-flavor symmetry is only broken by 
the electric charge operator as in Eq.(\ref{para2}).
We observe that in this limit the magnetic octupole 
moments are proportional to the baryon 
charge. 
\begin{table}[htb]
\begin{center}
\caption[M3 moments]{Magnetic octupole moments of decuplet baryons. 
Second column: SU(3) flavor symmetry limit ($r=1$).  
Third column: with flavor symmetry breaking ($r\ne 1$).} 
\begin{tabular}{ l | c |  c } & 
$\Omega_{B^*}(r=1)$  & $\Omega_{B^*}(r\ne 1)$   \\
\hline
$\Delta^{-}$     & $ -4C $	  & $-4C $           \\
$\Delta^{0}$     &     0     	  & 0                   \\
$\Delta^{+}$     & $4C $ 	  & $4C $            \\
$\Delta^{++}$    & $8C $  	  & $8C$            \\
\hline
 & &   \\
$\Sigma^{\ast -}$ & $-4C$ &      $-4C\,(1+r+r^2)/3 $      \\
$\Sigma^{\ast 0}$ & $0$       & $ -2C \,(2-r-r^2)/3$   \\    
$\Sigma^{\ast +}$ & $4C$  & $ -4C\,(1 -2r -2r^2)/3$  \\ 
\hline
  & & \\
$\Xi^{\ast -}$ & $-4C$     & $-4C\,(r + r^2 +r^3)/3$     \\
$\Xi^{\ast 0}$ &  $0$          & $-4C\,(r+r^2-2r^3)/3 $ \\ 
\hline  
  & & \\
$\Omega^-$ & $-4C$	  & $-4C\,r^3 $          \\  
\end{tabular} 
\label{octumom}
\end{center}
\end{table}
In order to estimate the degree of SU(3) flavor symmetry breaking 
beyond first order, we replace the spin-spin terms in Eq.(\ref{para2}) by 
expressions with a ``cubic'' quark mass dependence
$$
\sigma_{i} \sigma_{j} \rightarrow \sigma_{i} \sigma_{j}m_u^3/(m_i^2 m_j).
$$
This replacement is motivated by the flavor depen\-dence 
of the gluon exchange current diagram~\cite{Hen02}.  
Flavor symmetry breaking is then characterized by the ratio $r=m_u/m_s$
where $m_u$ and $m_s$ denote the up and strange quark masses.
This leads to analytic expressions for the magnetic octupole 
moments $\Omega_{B^*}$ containing terms up to third order in $r$
as shown in the third column of Table~\ref{octumom}.

Because the 10 diagonal octupole moments can be expressed 
in terms of only one constant $C$, there
must be 9 relations between them. Given the analytical expressions in 
Table~\ref{octumom} it is straightforward to verify 
that the following relations hold 
\setcounter{equation}{6}
\alpheqn
\begin{eqnarray}
\label{rel6a}
0 & = & \Omega_{\Delta^{-}} + \Omega_{\Delta^+}, \\
\label{rel6b}
0 & = & \Omega_{\Delta^{0}}, \\
\label{rel6c}
0 & = & 2\, \Omega_{\Delta^{-}} + \Omega_{\Delta^{++}}, \\
\label{rel6d}
0 & = & \Omega_{\Sigma^{* -}}- 
2\,\Omega_{\Sigma^{* 0}}+\Omega_{\Sigma^{* +}} , \\
\label{rel6e}
0 & = & 3 ( \Omega_{\Xi^{* -}} - \Omega_{\Sigma^{* -}} ) - 
(\Omega_{\Omega^{-}} - \Omega_{\Delta^{-}} ) \\
\label{rel6f}
0 & = &  (\Omega_{\Xi^{* 0}} + 2\, \Omega_{\Xi^{* -}})  + 
(\Omega_{\Sigma^{* +}}  -  \Omega_{\Sigma^{* -}}) \\
\label{rel6g}
0 & = & \frac{1}{3}(1+r+r^2) \Omega_{\Delta^+} + \Omega_{\Sigma^{* -}}, \\
\label{rel6h}
0 & = &  r \, \Omega_{\Sigma^{* -}} - \Omega_{\Xi^{* -}}, \\
\label{rel6i}
0 & = & r^3 \, \Omega_{\Delta^-} -\Omega_{\Omega^-}.
\end{eqnarray}
\reseteqn
The first six relations do not depend
on the flavor symmetry breaking parameter $r$ and hold irrespective 
of the order of SU(3) symmetry breaking.  
In fact, Eqs.(\ref{rel6a}-\ref{rel6d}) are already
a consequence of the assumed SU(2) isospin symmetry of strong interactions.
Eq.(\ref{rel6e}) is the octupole moment counterpart of  
the ``equal spacing rule'' for decuplet masses. 
Other combinations of the expressions in 
Table~\ref{octumom} can be written down if desirable.
\begin{table}[htb]
\begin{center}
\caption[M3 moments]{
Numerical values for magnetic octupole moments 
of decuplet baryons in [fm$^3$] 
using Table~\ref{octumom} with $C=-0.003$. Second column: SU(3) 
flavor symmetry limit ($r=1$). Third column: with SU(3) 
flavor symmetry breaking ($r=0.6$).}
\begin{tabular}{ l   r  r   }
                  &  $\Omega_{B^*}(r=1)$ &   $\Omega_{B^*}(r=0.6)$ \\[0.15cm] 
\hline
$\Delta^{-}$	  & 0.012            &   0.012       \\
$\Delta^{0}$	  & 0                &   0       \\
$\Delta^{+}$	  & -0.012           &  -0.012       \\
$\Delta^{++}$	  & -0.024           &  -0.024      \\
$\Sigma^{\ast -}$ &  0.012           &   0.008      \\
$\Sigma^{\ast 0}$ &  0               &   0.002      \\
$\Sigma^{\ast +}$ &  -0.012          &  -0.004     \\ 
$\Xi^{\ast -}$    &   0.012          &   0.005     \\
$\Xi^{\ast 0}$    &   0              &   0.002     \\
$\Omega^-$	  &  0.012           &   0.003     \\   
\end{tabular} 
\label{octumomnum}
\end{center}
\end{table}

To obtain an estimate for $\Omega_{\Delta^+}$ 
we use the pion cloud model~\cite{Hen01}
where the $\Delta^+$ wave function without 
bare $\Delta$ and for maximal spin projection 
is writtten as 
\be
\vert \Delta^+ J_z=\frac{3}{2}\rangle = 
\beta'\Bigl ( 
\sqrt{\frac{1}{3}} 
\vert n' \pi^+ \rangle
+ 
\sqrt{\frac{2}{3}} 
\vert p' \pi^0 \rangle \Bigr )
\vert \uparrow  Y^1_1  \rangle.
\ee
In this model the magnetic octupole moment operator is a product 
of a quadrupole operator in pion variables and a magnetic
moment operator in nucleon variables
\be 
\Omega_{\pi N} = \sqrt{\frac{16 \pi}{5}} 
\, r_{\pi}^2 \, Y^2_0({\bf r}_{\pi}) \, \, \mu_N \,\tau_z^N \, 
\sigma_z^N.
\ee
Here, the spin-isospin structure of $\Omega_{\pi N}$
is infered from the $\gamma \pi N$ and $\gamma \pi$ currents 
of the static pion-nucleon model~\cite{Hen62}.

With these expressions the $\Delta^+$ magnetic octupole moment is readily 
calculated~\cite{Hen01} 
\be
\Omega_{\Delta^+} = -\frac{2}{15} \, {\beta'}^{2}\, r_{\pi}^2 \
\mu_N = Q_{\Delta^+}\, \mu_{N} = r_n^2 \, \mu_N,
\ee
where $Q_{\Delta^+}$ is the $\Delta^+$ quadrupole moment 
and $r_n^2$ the neutron charge radius. With the experimental value 
of the latter and $\mu_N$ expressed in $[{\rm fm}]$  one gets
$\Omega_{\Delta^+} =-0.012\,\,{\rm fm^3}$.
The negative value of $\Omega$ implies that the magnetic moment 
distribution in the $\Delta^+$ is oblate and hence  
has the same geometric shape as the charge distribution.
Numerical values for other baryon octupole moments can now be obtained 
using Eq.(\ref{twothree}) and the expressions in Table~\ref{octumom}.
These are listed in Table~\ref{octumomnum}.
\vspace{-0.2cm}

\section{Summary} 
A general parameterization method based on SU(6) 
spin-flavor symmetry and its breaking has been used to calculate 
baryon octupole moments in a non-covariant quark model approach. 
We have constructed the  magnetic octupole moment operator 
in spin-flavor space and have shown that it contains only  
third order symmetry breaking three-quark currents.
We have then calculated analytical expressions for the  
octupole moments of decuplet baryons and derived 
certain relations between them.
 
To draw a first conclusion concerning the spatial 
shape of the magnetic moment distribution in baryons we have 
estimated the magnetic octupole moment of the $\Delta^+$ 
in the pion cloud model. Our result can be expressed as 
the product of the $\Delta^+$ quadrupole moment and the 
nuclear magneton.
This means that the magnetic moment distribution 
in the $\Delta^+$ is oblate and
hence has the same geometric shape as the charge distribution.
Numerical results have also been derived
for other decuplet baryons.

It would be interesting to calculate the magnetic octupole 
moments in other models in order to check whether our finding 
of an oblate magnetic moment distribution in decuplet baryons
can be confirmed.



\begin{thebibliography}{}
\bibitem{comment0}
If two of these had the same particle index, spin commutation relations 
would reduce them to a single Pauli matrix.
\end{thebibliography}
\end{document}